\documentclass[%
 reprint,
 amsmath,amssymb,
 aps,
]{revtex4-1}
\usepackage{hyperref}
\hypersetup{
	colorlinks=true,
	linkcolor=blue,
	filecolor=magneta,      
	urlcolor=blue,
}
\usepackage{graphicx}
\usepackage{dcolumn}
\usepackage{bm}
\usepackage{scalerel}

\usepackage{tikz}
\usetikzlibrary{svg.path}
\definecolor{orcidlogocol}{HTML}{A6CE39}
\tikzset{
	orcidlogo/.pic={
		\fill[orcidlogocol] svg{M256,128c0,70.7-57.3,128-128,128C57.3,256,0,198.7,0,128C0,57.3,57.3,0,128,0C198.7,0,256,57.3,256,128z};
		\fill[white] svg{M86.3,186.2H70.9V79.1h15.4v48.4V186.2z}
		svg{M108.9,79.1h41.6c39.6,0,57,28.3,57,53.6c0,27.5-21.5,53.6-56.8,53.6h-41.8V79.1z M124.3,172.4h24.5c34.9,0,42.9-26.5,42.9-39.7c0-21.5-13.7-39.7-43.7-39.7h-23.7V172.4z}
		svg{M88.7,56.8c0,5.5-4.5,10.1-10.1,10.1c-5.6,0-10.1-4.6-10.1-10.1c0-5.6,4.5-10.1,10.1-10.1C84.2,46.7,88.7,51.3,88.7,56.8z};
	}
}
\newcommand\orcidicon[1]{\href{https://orcid.org/#1}{\mbox{\scalerel*{
				\begin{tikzpicture}[yscale=-1,transform shape]
					\pic{orcidlogo};
				\end{tikzpicture}
			}{|}}}
}
\begin{document}
\preprint{APS/123-QED}
\title{Gravitational waves in neutrino plasma and NANOGrav signal}
\author{Arun Kumar Pandey \orcidicon{0000-0002-1334-043X}}
 \email[Email:]{\\arunp77@gmail.com,\\ arun\_pandey@prl.iitgn.ac.in}
\affiliation{Department of Physics and Astrophysics, University of Delhi, Delhi 110 007, India}
\date{\today}
\begin{abstract}
	{\centering
	\bf Abstract\par}
The recent finding of the gravitational wave (GW) signal by the NANOGrav collaboration in the nHZ frequency range has opened up the door for the existence of stochastic GWs. In the present work, we have argued that in a hot dense neutrino asymmetric plasma, GWs could be generated due to the instability caused by the finite difference in the number densities of the different species of the neutrinos. The generated GWs have amplitude and frequency in the sensitivity range of the NANOGrav observation. We have shown that the GWs generated by this mechanism could be one of the possible explanations for the observed NANOGrav signal. We have also discussed generation of GWs in an inhomogeneous cosmological neutrino plasma, where GWs are generated when neutrinos enter a free streaming regime. We show that the generated GWs in an inhomogeneous neutrino plasma cannot explain the observed NANOGrav signal. We have also calculated the lower bound on magnetic fields' strength using the NANOGrav signal and found that to explain the signal, the magnetic fields' strength should have atleast value $\sim 10^{-12}$ G at an Mpc length scale.
\end{abstract}
\maketitle
\section{\label{sec:level1}Introduction}
Albert Einstein first predicted gravitational waves (GWs) in 1916 based on his well-known general theory of relativity. Stochastic gravitational waves are the relic GWs from the early phases of the universe. The detection of stochastic GWs can give us a great insight into the early universe and hence into the high energy physics. These waves can be a direct probe of physics before the recombination era. Despite being so famous and important for several reasons, it could not be directly detected until 2016. When VIRGO-LIGO collaborations announced the first detection of the GWs, it originated due to the merger of two black holes of masses 29 and 36 solar masses. This successful detection of GWs has raised hopes of observing stochastic GWs by susceptible detectors in the future. At present, many ground and space-based experiments are active or are suggested to look for such a signal.  The B-mode polarization of Cosmic Microwave Background (CMB) probes GWs by its indirect effects on the CMB \cite{Zaldarriaga:1996xe, Kamionkowski:1996ks}. The ground based experiments for example LIGO (Laser Interferometer Gravitational-Wave Observatory) and advanced VIRGO \cite{Acernese:2015gua}, KAGRA \cite{Somiya_2012} and LIGO-India \footnote{http://www.gw.iucaa.in/ligo-india/, https://www.ligo-india.in/} have sensitivity at frequency $f\sim 10^{1\sim 3}$ Hz. However, the space based GW observations such as LISA \cite{AmaroSeoane:2012km, amaroseoane2017laser}, DECIGO \cite{Kawamura:2011zz}, BBO \cite{Crowder:2005nr} have best sensitivity at frequencies $f\sim$ mHz. The GWs with lower frequencies ($f\sim 10^{-9}$ Hz) are searched for by Pulsar timing arrays (PTA) such as EPTA \cite{Lentati:2015qwp, Arzoumanian_2018}, PPTA \cite{Verbiest:2016vj} and NANOGrav \cite{arzoumanian2020nanograv}. 

Recently, one of the PTA experiments, the NANOGrav collaboration, after analyzing the 12.5 years pulsar timing data, reported a signal of the nHz frequency, which might be strong evidence for the stochastic gravitational waves \cite{arzoumanian2020nanograv}. Various possible explanations of the observed signal have been given to date. One of the possible sources of the above signal can be of astrophysical origin, for example, mergers of supermassive black-hole binaries \cite{Rajagopal:1994zj, Jaffe:2002rt, Wyithe:2002ep}. Primordial black holes can generate stochastic GWs and are investigated in references \cite{Vaskonen:2020lbd, Bhattacharya:2020lhc, DeLuca:2020agl, DeLuca:2020agl, ding2020gravitational, xin2020multimessenger, Cai:2019bmk}. String theory-motivated models are also given to describe the observed flat spectrum of GW in the frequency band \cite{Ellis:2020ena, blasi2020nanograv, Buchmuller:2020lbh} (for earlier works see references \cite{Vilenkin:1981bx, Vachaspati:1985tv, Binetruy:2012ze, Gasperini_95, Lemoine_95, Ringeval:2005kr, Siemens:2006yp}). The gravitational waves sourced by magnetic fields and turbulence during the various phases of the early universe are discussed in references \cite{neronov2020nanograv,  Vagnozzi:2020gtf, Nakai:2020oit, Caprini:2010xv}. Some of the older works based on quantum fluctuation during inflation \cite{Rubakov:1982df, Giovannini:1999m, Sharma:2019jtb, Sharma:2021rot}, phase transitions \cite{Kamionkowski:1994km, Kosowsky:1992tw, Witten:1984we}, turbulent phenomenons \cite{Kosowsky:2002tm, Dolgov:2002gn, Kahniashvili:2005gr}, cosmic strings \cite{Vilenkin:1981bx} and the magnetic fields \cite{Anand:2018mgf, Pandey:2019tmo, Fujita:2020rdx} are the few important sources. In the present work, we have discussed the generation of stochastic GWs at the neutrino decoupling time. Authors of the reference \cite{Dolgov:2001nv} have addressed the generation of GWs when inhomogeneous neutrinos enter  the free streaming regime at the neutrino decoupling epoch. The non-linear interaction of neutrino flux with the collective plasma oscillations in a hot dense plasma causes instability and generates turbulence \cite{BINGHAM:1994rj, Tsytovich:1998vn}. This requires a net lepton number density, $n_\alpha(x) = n_{\nu_\alpha}(x) - n_{\nu_{\bar{\alpha}}}(x)$ (here $\alpha=e, \mu, \tau$, bar denotes the antiparticle) of the neutrinos species before the neutrino decoupling epoch at a length scale $L$, less than the Hubble horizon. The gradient of net lepton number density produces an electric current when the mean free path $\lambda_\nu$ of the neutrinos grows and becomes of the order of $L$ and generates magnetic fields \cite{Dolgov:2001nv, Boldyrev:2004sc}. These magnetic fields contribute to an anisotropic energy-momentum tensor and act as a source of gravitational waves. This mechanism works when we consider an inhomogeneous distribution of the lepton number density. However, when a parity-violating interaction of the neutrinos with the leptons is considered, it is shown that the necessity of an inhomogeneous distribution is no longer required \cite{Bhatt:2016hyi}. In this case, interactions modify the magneto-hydrodynamic equations \cite{Haas:2017jc} and contribute to an additional photon polarization \cite{Bhatt:2016hyi}. This additional contribution leads to a new kind of instability of the magneto-hydrodynamic modes and generates the magnetic fields at the cost of the homogeneous net lepton number density of neutrinos \cite{Pandey:2019tmo}. The total energy-momentum tensor contains anisotropic stress from the magnetic fields, and they act as a source of the primordial GWs. 

We have divided this paper in four sections. In section (\ref{sec-1-GWne}), we have discussed the generation of the magnetic fields and the gravitational waves at the neutrino decoupling epoch in two scenarios: due to parity odd interactions of the leptons with the neutrinos in a neutrino asymmetric plasma, and, due to inhomogeneous distribution of neutrinos at the time of neutrino decoupling epoch. Section (\ref{sec:resuldis}) contains the discussion of the models discussed in the previous sections in the context of NANOGrav signals and PTA, SKA observations. In the end, we have concluded the result of the present work in section (\ref{sec-1-concl}). In the present work, we have used Friedman-Robertson-Walker metric for the background space-time 
\begin{equation}
	ds^2=-a^2(\tau)d\tau^2+a^2(\tau)\delta_{ij} dx^i dx^j,
\end{equation}
where the conformal time and the coordinate are represented by $\tau$ and $x^i$ respectively and they are connected to the physical coordinate by the relations $d\tau =dt/a$ and $x=x_{\rm phy}/a$. Above metric is defined in such a way that the scale factor $a(\tau)$ has dimension of length. For a radiation dominated universe $a=1/T$ and the conformal time $\tau=\left(\frac{90}{8\pi^3 g_{\rm eff}}\right)^{1/2} \frac{M_{pl}}{T}$, $g_{\rm eff}$ is the effective relativistic degree of freedom at the epoch. We would also like to note here that, we have used natural unit system throughout the present work (for which $\hbar=c=k_B=1$). 
\section{Gravitational waves in a neutrino plasma}
\label{sec-1-GWne}
In this section, we discuss two cases of the generation of GWs, \textbf{i).} when parity-violating lepton-neutrino interactions are present, \textbf{ii).} when an inhomogeneous neutrino plasma enters a free streaming regime. Later a brief description of GW production by individual mechanism is given.
\subsection*{Case-i: Parity violating interactions of the neutrinos}
Parity violation in the context of electron-nucleon interactions ($e^{-}+N\rightarrow e^{-}+N$) is well studied. In various experiments, it has been shown that $e^{-}$ and $N$ (or in terms of electrons $e^{-}$ and quarks $q$) are coupled not only by electromagnetic interactions but they are also engaged in neutral weak coupling (for more details see table-1, for Neutral weak interactions in reference \cite{Commins:1980xv}). In a hot dense system (for example, in neutron stars), neutrinos behave abnormally and show asymmetry in the particles' number densities over antiparticles. This abnormal behavior of neutrinos is one of the prominent cosmological \textit{puzzle}. The beyond Standard Model (BSM) framework is one of the possible ways to understand this \textit{puzzle}. The interaction of the neutrinos $\nu$ with the leptons $l$ is given by the effective current-current Lagrangian \cite{Commins:1980xv, Giunti:2007ry}
\begin{equation}
	\mathcal{L}_{eff} =-\sqrt{2}G_F \sum_{\nu_e, \nu_\mu, \nu_\tau}{J_{\nu}}_\alpha \, J^\alpha_l\, ,
\end{equation} 
where $G_F\approx 1.17\times 10^{-5}$ GeV$^{-2}$ is the Fermi constant and ${J_{\nu}}_\alpha$ and $J^\alpha_l$ are defined as axial neutrino currents and vector lepton current. The summation in the above equation is to consider all neutrinos generations. The ensemble average over the neutrino current gives $\langle{J_{\nu}}_\alpha \rangle \approx (n_{\nu_\alpha} - n_{\bar{\nu}_\alpha})$. This additional current $j_\nu^\beta$ can be written in terms of polarization tensors $\Sigma^{\alpha\beta}/$ as $j_\nu^\beta=\Sigma^{\alpha\beta}A_\alpha$. Here the polarization tensor $\Sigma^{\alpha\beta}$ contains three terms, longitudinal, transverse and parity odd term denoted by $\Sigma^{\alpha\beta}_L$, $\Sigma^{\alpha\beta}_T$ and $\Sigma^{\alpha\beta}_A$ respectively. The odd parity term in the current expression leads to instability in the neutrino plasma, leading to turbulence \cite{Bhatt:2016hyi, Akamatsu:2013yy, Dvornikov:2013bca}. The total three current in this case is given by
\begin{equation}
	\label{sec:currnu}
	j_\nu^i \approx \sigma \, E^i- \Sigma_2 B^i- \Sigma_\nu \omega^i,
\end{equation}
where, $\Sigma_\nu$ is given by $\Sigma_\nu \propto T^2$ for $\mu \ll T$ and $\Sigma_\nu \propto \Delta \mu^2$ for $\mu \gg T$ (where $\mu$ is the chemical potential). $\Sigma_2$ represents the neutrino asymmetry and it is given in reference \cite{Pandey:2019tmo} (see equation 14). In this equation \eqref{sec:currnu}, the first term represents the ohmic current, the second and third terms come only for the neutrino asymmetric neutrino plasma.  Here $\sigma$, $\Delta \mu$ and $\alpha$ represent the conductivity, asymmetry in the number density of the neutrinos species and the electromagnetic (EM) coupling constant respectively. The symbols ${\bf E}$, ${\bf B}$ and $\boldsymbol{\omega}=\nabla\times {\bf V}$ are the electric, magnetic and vorticity three vectors respectively (here ${\bf V}$ is the velocity vector). Last term in the above equation leads to a term proportional to $\nabla \times \boldsymbol{\omega}$ in the magnetic induction equation, which produces a sufficiently strong magnetic fields. In reference \cite{Pandey:2019tmo}, authors have shown that, magnetic fields are generated at the cost of this turbulent kinetic energy. Generation of the magnetic field in the present scenario is given by the following equations
\begin{eqnarray}
	\frac{\partial E_B}{\partial \tau}  = \left( -\frac{2k^2}{\sigma} + \frac{\Sigma_2}{\sigma}\right) E_B+\frac{2\Sigma_\nu^2 k^4}{\sigma^2}(\tau-\tau_*)E_{\rm v}
\end{eqnarray}
where $E_B$, and $E_{\rm v}$ represents the magnetic energy and the turbulent energy density respectively. $\tau_*$ is the initial time at which the turbulence is generated and hence the magnetic fields. When sufficiently strong magnetic fields are generated, first term on the right hand side dominates over second term and hence solution can be written as: $E_B\propto {\rm Exp}\left[-2\tau k/\sigma(k- \Sigma_2)\right]$. For the wave number $k\leq \Sigma_2$, magnetic modes will grow exponentially. However for wave number $k>\Sigma_2$, modes will damp.
\subsection*{Case-ii: when inhomogeneous neutrinos enter the free streaming regime}
The production of GW, in this case, is briefly described in reference \cite{Dolgov:2001dd}. This mechanism works at a neutrino decoupling epoch. At this epoch, the Hubble horizon was significantly larger than that of EW and QCD phase transitions. This mechanism works when there are net inhomogeneous lepton number density ($\Delta n_\alpha(x) = n_{\nu_\alpha}(x) - n_{\nu_{\bar{\alpha}}}(x)$) of one, or more species of the neutrinos before neutrino decoupling epoch at certain length scale ($L< H^{-1}$). At the epoch of neutrino decoupling, elastic scattering of the neutrinos to the electrons and positrons creates turbulence, and hence a vortical motion in the plasma. As a result, a gradient of net lepton number density produces an electric current, and hence magnetic fields, when the mean free path $\lambda_\nu$ of the neutrinos grow and become of the order of $L$ \cite{Dolgov:2001nv, Boldyrev:2004sc}. This can be seen through the following equation \cite{Dolgov:2001dd}
\begin{equation}
	\frac{\partial \mathcal{K}_\nu}{\partial \tau}\propto \frac{\delta n_\nu}{n_\nu},
\end{equation}
where $\mathcal{K}_\nu$ is the specific momentum flux at a scale $L$. The term on the right hand side produces a non-zero vorticity, i.e $\nabla \times {\bf V}\neq 0$ and hence magnetic fields, which can be understood by the Biermann battery equation $\frac{\partial {\bf B}}{\partial \tau}\propto (\nabla p \times \nabla \rho)$ (here $p$ and $\rho$ are the pressure and the number density respectively).
\subsection*{Gravitational wave production}
These magnetic fields can contribute to anisotropic stress to the total energy-momentum tensor $T_{ij}$. The transverse traceless (TT) part of energy-momentum tensor $T_{ij}$ can then source the metric perturbation and generates the gravitational waves. The tensor metric perturbations are defined by the metric 
\begin{eqnarray}
	ds^2=-a^2(\tau) d\tau^2+a^2(\tau)(\delta_{ij}+2h_{ij})dx^i dx^j
\end{eqnarray}
In terms of comoving coordinates and time, the evolution equation of $h_{ij}$ are
\begin{equation}
	\label{eq:GW-eq}
	h''_{ij}(\tau, k)+2\mathcal{H}h'_{ij}(\tau, k)+k^2 h_{ij}(\tau, k)=16 \pi G\, \Pi^{TT}_{ij}(\tau, k),
\end{equation}
where $\mathcal{H}=\frac{1}{a}\frac{da}{d\tau}$ is the comoving Hubble parameter. For the radiation dominated and matter dominated, $\mathcal{H}$ is given by $1/\tau$ and $2/\tau$ respectively. On the right hand side of equation (\ref{eq:GW-eq}), $\Pi_{ij}^{TT}$ is the transverse traceless component of the energy momentum tensor contributed by the magnetic fields, generated in a neutrino plasma and it is given by $\Pi_{ij}^{TT}({\bf k})=\mathcal{P}_{ikjl}({\bf k})\,T_{kl}(k)$. Here $T_{ij}$ is given as
\begin{equation}
	\label{eq:TmynuB}
	T_{ij}=\frac{1}{a^2}\left(B_i B_j -\frac{1}{2}\delta_{ij} B^2\right)
\end{equation}
The energy density of the GW is defined as  \cite{Isaacson:1968ar}:
\begin{eqnarray}
	\rho_{\rm gw}(\tau, {\bf x}) =\frac{1}{32\pi }\langle h'_{ab}(\tau, {\bf x}) h'^{ab}(\tau, {\bf x}) \rangle 
\end{eqnarray}
We define the energy density power spectrum in Fourier space as
\begin{equation}
	\label{eq:omegafk}
	\frac{d\Omega_{\rm gw}}{d {\rm log} k} =\frac{k^3}{2(2\pi)^3G\rho_c a^2} |h'|^2=\frac{k^5}{2(2\pi)^3G\rho_c a^2} \left|\frac{dh}{dx}\right|^2
\end{equation}
where $\rho_c= 3H_0^2/8\pi G$ is the present day critical density and $x=k\tau$. The GW energy power spectrum is defined as
\begin{equation}
	\langle h'_{ij}(\tau, {\bf p}) h'^*_{ij}(\tau, {\bf q}) \rangle	=(2\pi)^3\delta({\bf p}-{\bf q})\, |h'(\tau, p)|^2.
\end{equation}
The generated GW will decay only through the expansion of the universe and scales as $1/a^4$, when the considered wavelength is inside the horizon (i.e $x\gg 1$). At such a scale, GW energy density power spectrum at present is given by \cite{Pandey:2019tmo}
\begin{eqnarray} \label{eq:om*}
	\left.\frac{d\Omega_{\rm gw}}{d \ln k}\right\vert_0  = \left.\frac{\Omega_{\rm GW}}{d \ln {k}}\right\vert_* \left(\frac{g_{s_0}}{g_{s_*}}\right)^{4/3}\left(\frac{T_0}{T_*}\right)^4 \frac{\rho_{c,*}}{\rho_{c,0}}\,,
\end{eqnarray}
where $g_s$ is the degree of freedom, $T$ is the temperature, subscript `0' and `*' shows that values at present-day and the generation time respectively. To find this expression, we need to find the solution of equation (\ref{eq:GW-eq}). It has been shown in reference \cite{Pandey:2019tmo} that, in the case of neutrinos in a hot dense plasma, generated GW energy density power spectrum at present is given by
\begin{eqnarray}
	\label{eq:GW-spec-0}
	\left.\frac{d\Omega_{\rm gw}}{d \ln k}\right\vert_0 & = &   \frac{64\pi M_{\rm st}^2 k^3}{9(2\pi)^6}\,  \left(\frac{g_{s0}}{g_s*}\right)^{4/3} \, \left(\frac{T_0}{T_*}\right)^2\, \nonumber \\
	& &\left(\frac{T_0}{H_0}\right)^2\,\left[\ln(x_{*})\frac{\partial }{\partial \tau}\,\left.\left(\frac{\sin x}{x}\right)\right\vert_{\tau=\tau_*}\right]^2 \, f(k)\, ,
\end{eqnarray}
where $M_{\rm st}=\left(\frac{90}{8\pi^3 g_{\rm geff}}\right)^{1/2}M_{\rm pl}$. Here $f(k)$ is given by the relation
\begin{eqnarray}
	\label{eq:mag-fk}
	f(k) &=& \frac{1}{4}\frac{1}{(4\pi)^2} \int d^3p\,[(1+\gamma^2)(1+\beta^2)\,S(p)S(k-p) \nonumber \\
	&+&  4\,\gamma\,\beta\,A(p)\, A(k-p) \, ].
\end{eqnarray}
In above equations, $\gamma= \hat{{\bf k}}\cdot \hat{{\bf p}}$ and $\beta= \hat{{\bf k}}\cdot(\widehat{{\bf k}-{\bf p}})\,$. $S(k)$ and $A(k)$ are related to the magnetic energy and helical magnetic energy density of the magnetic fields and they are defined as: $\Omega_B(k)=2\pi k^3 \, S(k)/\rho_c$ and $\Omega_H(k)=2\pi k^2 \, A(k)/\rho_c$. From the equation (\ref{eq:GW-spec-0}) it is thus clear that the energy density spectrum of GWs are mainly governed by the behavior of $f(k)$ at different scales, defined in equation (\ref{eq:mag-fk}).
\section{Result and Discussion}
\label{sec:resuldis}
In this section, we interpret the NANOGrav signal as a primordial GWs background generated by the magnetic fields in a hot dense parity odd neutrino plasma. Later we discuss the statistical properties of the GWs induced in a neutrino asymmetric plasma and calculate the favored slop in current context. We have also calculated the strength of the magnetic fields from the observed NANOGrav signal.
\begin{figure}
	\centering
	\includegraphics[width=3.2in]{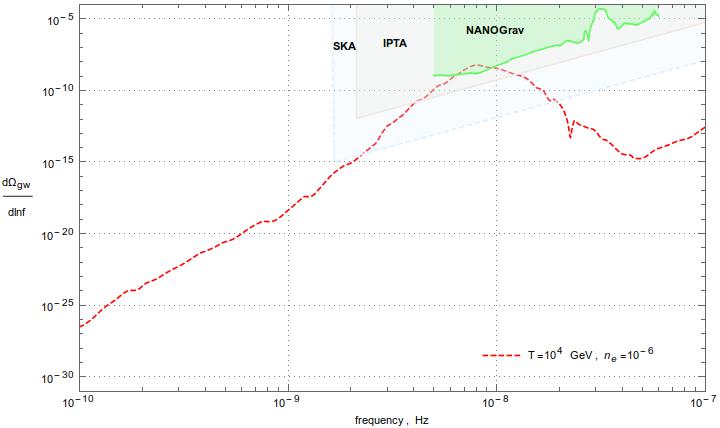}
	\caption{The light green shaded area is the sensitivity range of the NANOGrav observation. The light orange range represents the sensitivity range of the IPTA experiment and the light blue shaded range is the sensitivity of the SKA. The dashed red lines show the GW power spectrum for the GWs generated by the magnetic fields, which have sourced by the magnetic fields, produced in a hot dense neutrino plasma (electron density of $10^{-6}$). Here we have chosen the plasma electrical conductivity of $\sigma\sim 100$.}
	\label{fig:comparitive-plot}
\end{figure} 
\subsection{\label{sec-1-GWNANOGrav}NANOGrav signal and neutrino induced GW signature}
Figure (\ref{fig:comparitive-plot}) shows the plots of the present study (red-dashed line) along with the sensitivities of the NANOGrav (light-green shaded region), IPTA (light-orange) and the SKA (light-blue) collaborations. In the present study, we show that the generated magnetic fields in a hot dense neutrino plasma act as a source of the GWs (see equation (\ref{eq:GW-eq}) and (\ref{eq:TmynuB})).  In figure (\ref{fig:comparitive-plot}), it is clear that the amplitude and the frequency of the GWs (red-dashed line) generated by magnetic fields at temperature $T\sim 10^4$ GeV in a neutrino asymmetric plasma lie in the allowed range of the NANOGrav experiment at a frequency ($f\sim 10^{-8}$ Hz). At lower temperatures, the spectrums' peak shifted towards the lower frequency, but amplitudes are so small that they are out of reach of the NANOGrav sensitivity. In a previous work \cite{Pandey:2019tmo}, it has been shown that the GWs originated at lower temperatures ($T< 10^4$ GeV) can be detected in the IPTA and SKA observations. Therefore we, believe that one of the possible explanations for the detected signal by NANOGrav collaborations is the GWs generated in a neutrino plasma above neutrino decoupling, where parity odd interactions of the neutrinos with the leptons are dominant. In this case, we need not have to consider the inhomogeneous distribution of the neutrinos in the plasma. In the case of inhomogeneous neutrino density, the produced GWs have energy density $\Omega_{\rm gw}h^2\simeq 9.5 \times 10^{-7}\lambda_0^{3/4} v^6 (x_*)$ and frequency $f_0\simeq 10^{-7} \lambda_0^{-1}$ (here $\lambda_0$ is the comoving wavelength, $f_0$ is corresponding frequency and $v$ represents the velocity field) \cite{Dolgov:2001nv}. We have found that, although amplitude ($\sim 10^{-10}$) lies in the NANOGrav lower limit, frequency ( $f\sim 10^{-6}$ Hz) is beyond the reach of the experiment.  Therefore, the produced GWs can not explain the observed GWs by the NANOGrav collaboration. Instead, these GWs could be detected by space-based experiments (eLISA). 

Earlier, various lepton asymmetric and phase transition models were given to explain the observed magnetic fields and hence GWs. For example, in reference \cite{Anand:2018mgf} (for similar work based on baryogenesis and leptogenesis, see references \cite{Beniwal:2018hyi, Xie:2020bkl}), it has been shown that at a temperature above EW phase transitions (T$\geq 100$ GeV), due to the chiral asymmetry of the electrons, strong magnetic fields are generated and these fields later act as a source for the GWs. However, the frequency of the GWs generated due to the magnetic fields at temperature T$\sim 10^4$ GeV in a chiral asymmetric fluid is of the order of $\sim 10^{-1}$ Hz. Therefore, these GWs can not explain the observed NANOGrav signal. In EW phase transition models, for example in reference \cite{Kamionkowski:1994km}, produced GWs can not explain NANOGrav signals as the frequency of these GWs are of few mHz. In these models, the source of the gravity waves is colliding bubbles and hydrodynamic turbulence at the cosmological phase transitions.  In a more recent work based on the QCD phase transition \cite{neronov2020nanograv} (see also \cite{Caprini:2010xv}), authors have described the observed NANOGrav signal as a product of the magnetohydrodynamic (MHD) turbulence at the phase transition. Gravitational-wave signatures resulting from the strong first-order phase transition due to the presence of the Higgs doublet have been discussed in the reference \cite{Barman:2019oda}. However, the frequency ($10^{-2}-10^{-3}$ Hz) of these GWs is much higher than the frequency of the observed GW by NANOGrav collaboration and hence again cannot explain the NANOGrav signal. A similar situation of lepton number asymmetry can arise in the case of Quark-Gluon Plasma (QGP) at temperature $T\sim 100$ MeV. In the case of the merger of neutron stars in a binary system, the observational signature of the gravitational waves is discussed in reference \cite{Abbott:2017gw2}. It is believed that quarks and gluons are the major constituents at such a high temperature in the core of the neutron stars. Such mergers represent potential sites for a phase transition from a confined hadronic matter to deconfined quark matter.  In a fully general-hydrodynamic simulation, it is shown that a similar GWs signature from the merger of neutron stars GW170817 (LIGO collaboration \cite{Abbott:2017gw2}) can be obtained in the case of QGP phase transition \cite{Weih:2020hr}. The obtained frequency of these GWs are in the sensitivity range of LIGO and hence cannot explain the NANOGrav signal. Therefore, we believe that of all possible models based on the phase transitions, chiral asymmetric models above EW phase transitions, neutrino asymmetric models of generation of GWs is one of the suitable models to explain the NANOGrav signal.
\subsection{Statistical properties and power law background}
From various theoretical magnetohydrodynamic models, it is expected that the power spectrum of the stochastic GW background to be a broken power-law $f^\beta$. In a super Horizon frequency range, where frequency $ f <f_H=aH$, the slop $\beta=3$. However, around source frequency, $0 < \beta \leq 3$ \cite{Caprini:2009fx} and at frequencies $f\geq f_*$ (here $f_*$ is the source frequency), the slop $\beta <0$ \cite{Niksa:2018ofa}. Normally slop $\beta$ depends on the initial conditions of magnetic fields, type of MHD turbulence and its temporal evolution and the decorrelation time. The characteristic strain spectrum $h_c(f)$ describes the GW background in the experiments and it is normally expressed as a function of dimensionless amplitude $A$ at a reference frequency $f_{\rm yr}=1/{\rm yr}\sim 10^{-8}$ Hz  (inverse of time in year)
\begin{equation}
	\label{eq:hcf}
	h_c(f)= A \left(\frac{f}{f_{\rm yr}}\right)^\alpha.
\end{equation}
In above equation, the parameter $\alpha$ is the slop of the GW strain. The scaling of the $\Omega_{\rm GW}$ is interpreted as the frequency dependence at the peak of the the GW power spectrum. To understand the observed stochastic GW background on a detectors, we need to compare the theoretical model to the fitted power law given in equation (\ref{eq:hcf}). In a transverse traceless gauge 
\begin{eqnarray}
	\langle h_{ij}^{\rm TT}(t)\, h_{ij}^{\rm TT}(t)\rangle = 2\int_{f=\infty}^{f=f}\, d \,{\rm log}f\, \, h_c^2(f)\, ,
\end{eqnarray}
where the angular brackets denote the ensemble average for the stochastic GW background. The factor $2$ on the right hand side in above equation is motivated by the fact that, in an unpolarized background, the left hand side is made up of two contributions, $\langle h_+^*h_+\rangle$ and $\langle h_-^*h_-\rangle$. The GW power spectrum is given by 
\begin{equation}
	\frac{d\rho_{\rm GW}(f)}{d {\rm log}f}=\frac{\pi c^2}{4\, G}\, f^2\, h_c^2(f).
\end{equation}
Therefore, power spectrum of the GW, using equation \eqref{eq:omegafk} can be expressed as
\begin{eqnarray}
	\label{eq:omegafk-2}
	\frac{d \Omega_{\rm GW}}{d {\rm log}\, f} & =&\frac{1}{\rho_c}\frac{d\rho_{\rm GW}(f)}{d {\rm log}f} =\frac{\pi c^2}{4\, G\rho_c}\, f^2\, h_c^2(f)\nonumber \\
	& =& \frac{\pi c^2}{4\, G\rho_c}\, f^2\, A^2 \left(\frac{f}{f_{\rm yr}}\right)^{2\alpha}
\end{eqnarray}
Now comparing equations (\ref{eq:omegafk}) and (\ref{eq:omegafk-2}), in a large scale limits ($k\tau \ll1$):
\begin{eqnarray}
	A^2 & = &  \frac{128 M_{\rm st}^4 G \rho_c f_{\rm yr}^{3}}{81 \pi H_0^2}\, \left(\frac{g_{s0}}{g_s*}\right)^{4/3} \, \left(\frac{T_0}{T_*}\right)^4\, \nonumber \\
	& \times& \left[{\rm log}\left(\frac{2\pi f_* M_{\rm st}}{T_*^2}\right)\right]^2\, B_0^4\, ,
	\\
	\alpha & = & \frac{3}{2}\, .
\end{eqnarray}
Here we have compared the two equations at the peak of the GW spectrum for the superhorizon GW modes after considering that near peak, $f(k)\approx B_0^4$ (where $B_0$ is the present day large scale magnetic field strength). Therefore, for a GW produced in a neutrino asymmetric plasma at neutrino decoupling epoch, $\alpha=3/2$ at the peak of the GW power spectrum \cite{arzoumanian2020nanograv, neronov2020nanograv, Caprini:2019egz, Niksa:2018ofa}. The slop calculated here is well within the 2$\sigma$ bounds obtained for the slop of the power law spectrum of the GWs by the NANOGrav collaborations (see figure 1 in reference \cite{arzoumanian2020nanograv}).
\subsection{Favored strength of magnetic fields}
From equation (\ref{eq:TmynuB}), we can write the transverse traceless part of the total energy-momentum tensor as: $\Pi^{TT}\sim B^2/2$. In  equilibrium (i.e when $h_{ij}=h_{ij}'=0$), tensor perturbations $h\sim 16\pi G \Pi^{TT}/k^2\simeq 8\pi G B^2/ak^2$. The comoving energy density spectrum per logarithmic scale, in terms of the magnetic field, can be expressed as $\Omega_{\rm gw}=k^2 h^2/(32\pi G\, \rho_{c})\propto f(k)$. It is thus apparent that the spectrum of the GW depends on the nature of the function $f(k)$ and hence on the magnetic field origin method. The strength of the magnetic fields, induced in a neutrino asymmetric plasma in a hot dense plasma at the time of neutrino decoupling, could be constrained by considering the fact that the generated GWs will have amplitude atleast in the NANOGrav sensitivity range. Which means that $\Omega_{\rm GW, 0}\geq k^2 h_{\rm min}^2/(32\pi G\, \rho_{c})$ and hence 
\begin{equation}
	B_{\rm min}^2 = \sqrt{\frac{\Omega_{\rm GW, 0} \rho_c}{2 G} \left(\frac{2\pi f}{c}\right)^2 }
\end{equation}
It is thus obvious that to explain NANOGrav signal at frequency $f\sim 10^{-8}$ Hz, strength of the magnetic fields generated at temperature T$\sim 10^4$ GeV, should have  a minimum value of the order of $1.3\times 10^{-12}$ G at a coherence scale of $1$ Mpc length scale at present.

%
%
%
\section{Conclusion}
\label{sec-1-concl}
In conclusion, we have found that the GWs produced in a homogeneous neutrino plasma can have amplitude and the frequency in the sensitivity range of the NANOGrav experiment if these GWs were generated much above the neutrino decoupling epoch and there are parity odd interactions between the neutrinos and the leptons. However, GWs generated in an inhomogeneous neutrino plasma, sourced by the magnetic fields, cannot explain the observed NANOGrav signal. It is thus clear from the present work that apart from the well-studied mechanism of the generation of the primordial GWs by various phase transitions, inflation, or some turbulent phenomena in the early universe, the proposed mechanism in the present study is also one of the possible explanations for the observed signal.
\begin{acknowledgments}
	A.K.P. is financially supported by the Dr. D.S. Kothari Post-Doctoral Fellowship, under the Grant No. DSKPDF Ref. No. F.4-2/2006 (BSR)/PH /18-19/0070. A. K. P would also likes to thanks Prof. T. R. Seshadri and Dr. Sampurn Anand for the useful discussions and the comments during the work.
\end{acknowledgments}
\bibliographystyle{apsrev4-1}
\bibliography{NANOGrav-neu}
\end{document}